\documentclass[aps, prl, twocolumn,superscriptaddress,amsmath,amssymb,floatfix]{revtex4-1} 

\usepackage{graphicx}
\usepackage[usenames]{color} 
\usepackage{comment} 
\usepackage{bm}
\usepackage{amsmath}
\newcounter{abc}

\newcommand{\br}{{\bf r}} 
\newcommand{\bu}{{\bf u}} 

\newcommand{\be}{\begin{equation}} 
\newcommand{\ee}{\end{equation}}
\newcommand{\bea}{\begin{eqnarray}} 
\newcommand{\eea}{\end{eqnarray}}

\begin{document}

\title{Quantifying the effects of neglecting many-body interactions in coarse-grained models of complex fluids}

\author{Douglas J. Ashton}  
\author{Nigel B. Wilding} 
\affiliation{Department of Physics, University of Bath, Bath BA2 7AY,
United Kingdom} 

\begin{abstract}

We describe a general simulation scheme for assessing the
thermodynamic consequences of neglecting many-body effects in
coarse-grained models of complex fluids.  The method exploits the fact
that the asymptote of a simple-to-measure structural function provides
direct estimates of virial coefficients.  Comparing the virial
coefficients of an atomistically detailed system with those of a coarse-grained
version described by pair potentials, permits the role of many-body
effects to be quantified. The approach is applied to two models: (i) a
size-asymmetrical colloid-polymer mixture, and (ii) a solution of star
polymers. In the latter case, coarse-graining to an effective fluid
described by pair potentials is found to neglect important aspects of the true
behaviour.

\end{abstract}

\maketitle

Many-body forces occur when the net interaction between two particles
is not simply pairwise additive, but depends on the presence of other
particles. They appear in a wide range of physical systems including
dense phases of noble gases~\cite{Egelstaff:1987kl}, molecular
systems~\cite{Elrod:1994qa}, nuclear matter~\cite{Zuo:2013fu},
superconductors ~\cite{Carlstrom:2011dz} and complex fluids such as
polymers~\cite{DAdamo:2012sf}, lipid
membranes~\cite{Schmid:2006,Lyubartsev:2011xy} and colloidal
dispersions~\cite{Brunner:2004kx,Merrill:2010vn,Forsman:2012uq,Mattos:2013fk}.
In seeking to make theoretical and computational progress with such
systems one often attempts to simplify matters by ``coarse-graining''
ie. integrating over the degrees of freedom on small length or times
scales.  This leads to a description of the system in term of an
effective Hamiltonian describing the interactions among the remaining
degrees of freedom.  These interactions are inherently many-body in
character, even if the original system involves only pairwise
interactions.

\begin{figure}[h]
\includegraphics[type=pdf,ext=.pdf,read=.pdf,width=0.8\columnwidth,clip=true]{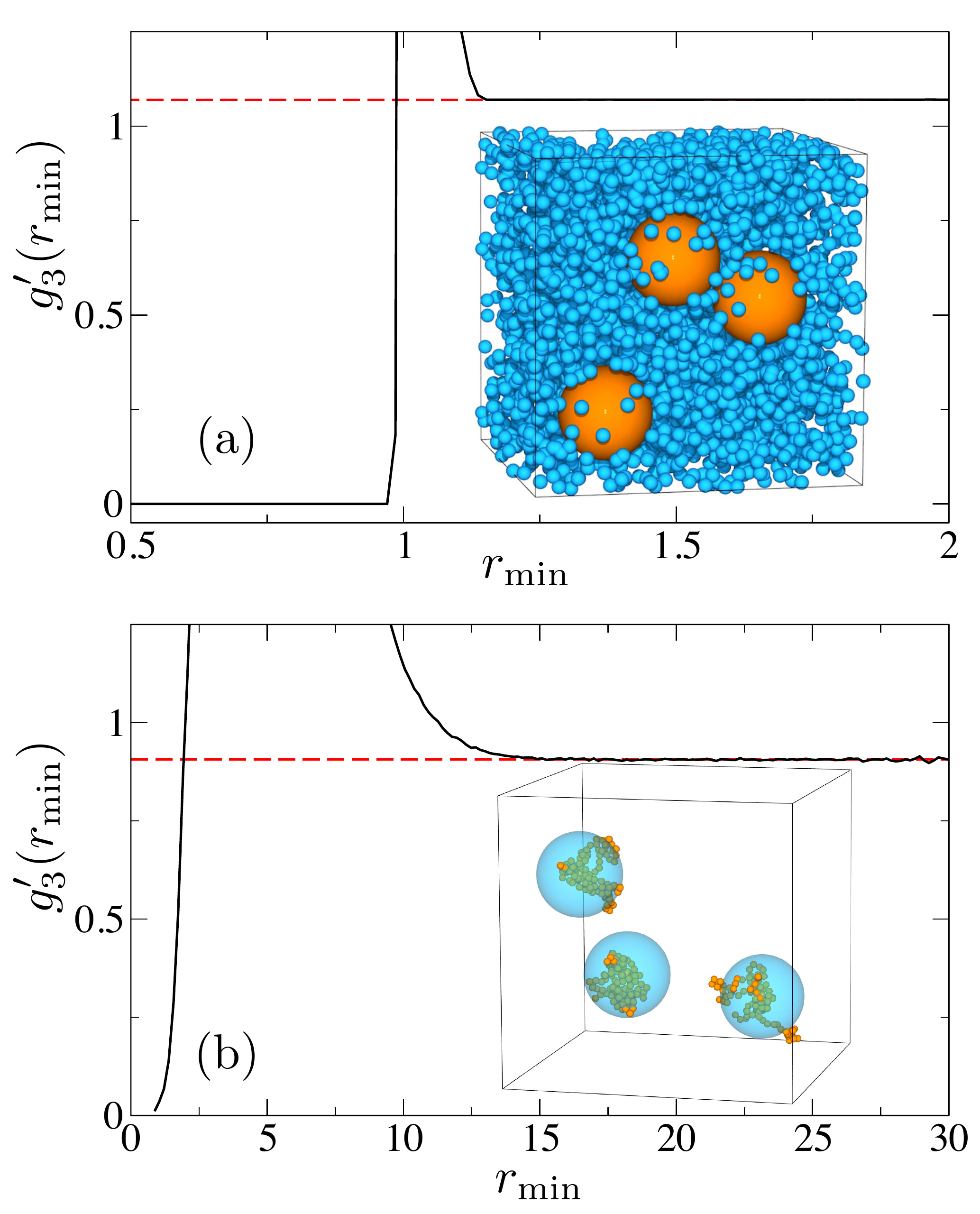}
\caption{(Color online). {\bf (a)} Snapshot of a highly size-asymmetrical mixture of
 spheres. The effective one-component model is realized by
 integrating out the small sphere degrees of freedom. {\bf (b)} A
 snapshot of three star polymers.
The big spheres represent a coarse grained model in which each polymer
is replaced by a single effective particle. In both (a) and (b) simulation measurements are also shown for
the structural quantity $g_3^\prime(r_{\rm min})$ discussed in the text, whose asymptote (dashed) yields information on the third virial coefficient.}
 \label{fig:illustr}
 \end{figure}

To appreciate how many-body interactions arise in coarse-grained (CG)
representations of complex fluids, consider the case of colloids
dispersed in a sea of much smaller polymers. This system is commonly
modelled as a highly size-asymmetrical mixture of spheres as shown in
the simulation snapshot of Fig.~\ref{fig:illustr}(a). However, since 
dealing with components of disparate sizes is theoretically and
computationally problematic, one typically seeks to integrate out the
polymer degrees of freedom to yield an effective one-component
model. But the colloidal interactions arise from the modulation
of the polymer density distribution by {\em all} the colloids, and
consequently, the effective one-component description is many-body in
form.  A second example is shown in Fig.~\ref{fig:illustr}(b) which
depicts three star polymers in solution.  A common CG model replaces
each star by a single effective particle. However, the net
interaction between two polymers depends on the proximity of a third,
and hence the effective Hamiltonian has a many-body
character~\cite{Ferber:2000fk}.

Computer simulation is a powerful route for designing CG
models for complex fluids, which is currently receiving considerable
attention. Indeed, in principle it can be used to determine a many-body
potential for the CG coordinates which is consistent with the
underlying atomistic model~\cite{Noid:2008ai,Noid:2008hc}. But
implementing such approaches is far from trivial and it is still the
norm that CG descriptions neglect some or all many-body effects~\cite{Muller-Plathe:2002qv,Nielsen:2004jk,Voth:2008,Rudzinski:2012fk}. Thus
for polymers one might replace each chain by a set of one or more
`blobs' which interact via a pair potential such as the potential of
mean force~\cite{Capone2008}. Similarly for a colloid-polymer mixture one typically
describes the colloidal interactions via the depletion pair
potential~\cite{Lekkerkerker:2011}. In view of this, it is patently important to be able to
assess the likely implications of the pair potential approximation for
the thermodynamics of the CG system. To date, though, there has been
little in the way of systematic methods for doing so.

In this Letter we introduce a widely applicable tool for comparing
some basic thermodynamic properties of an atomistically detailed
system with those of its CG representation.  Our approach is based on
calculations of the virial coefficients $B_n$. These are pertinent
because they provide a systematic expansion of the properties of a
system in terms of many-body interactions: $B_2$ depends only on pair
interactions, while $B_3$ depends on two- and three-body interactions,
etc. Comparison of virial coefficients for the atomistically detailed
and CG models provide a measure of the extent to which the
thermodynamics of the models agree.

Previously there has been no method capable of directly determining
virial coefficients for complex fluids.  The novel feature of our 
approach is that it is based on exploiting {\em finite-size
  effects}. Consider a simulation box of volume $V$ containing $N$
interacting molecules in thermal equilibrium at inverse temperature
$\beta=(k_BT)^{-1}$. For each of the $N$ molecules we tag an arbitrary atomic
site and label its position vector $\br_i$, with $i=1\ldots N$.  The
position vectors of the remaining $m$ atoms in each molecule we write
as $\br_{i,j}=\br_i+\bu_{i,j}, j=1\ldots m$, with $\bu_{i,j}$ the
displacement of atom $j$ on molecule $i$ from the tagged atom
$\br_i$. Accordingly a molecular configuration can be specified via a
list of the $N$ tagged and the $M=Nm$ non-tagged coordinates,
$\br^N,\bu^M$. The corresponding Boltzmann probability is
\be
P_N(\br^N,\bu^M)=\frac{ e^{-\beta U(\br^N,\bu^M)}} {Z_N} \:,
\label{eq:boltzmann}
\ee
where $U(\br^N,\bu^M)$ is the full interaction potential containing both intra and intermolecular terms and
\be
Z_N=\int e^{-\beta U(\br^N,\bu^M)}d\br^Nd\bu^M
\ee
is the $N$-molecule configurational integral.

Now define 
\be
\tilde{g}_N(\br^N,\bu^M) \equiv \frac{P_N(\br^N,\bu^M)}{P_N^{\rm ig}(\br^N)}= V^N \frac{ e^{-\beta U(\br^N,\bu^M)}}{Z_N}\:,
\label{eq:main}
\ee
where $P_N^{\rm ig}(\br^N)=V^{-N}$ is the probability of finding (within the same volume) a set of $N$ structureless ideal gas particles
in the same configuration as the tagged sites. We shall focus on the low density limit of
$\tilde{g}_N(\br^N,\bu^N)$, corresponding to $|\br_k-\br_l|\to\infty,
\:\;\forall\: k,l$. In this regime the molecules are non-interacting, so we
can integrate out the internal molecular degrees of freedom
(associated with the $\bu_{i,j}$) to obtain the asymptotic value

\be
f_N(V)\equiv\lim_{|\br_k-\br_l|\to\infty}\tilde{g}_N(\br^N)=\frac{(\Omega V)^N}{Z_N}=\frac{Z_1^N}{Z_N}\:,
\label{eq:limit}
\ee
where $\Omega$ is the integral over the internal degrees of freedom of
a single molecule and
$Z_1$ is the corresponding configurational integral.

The quantity $f_N(V)=Z_1^N/Z_N$ is central because it permits a direct
calculation of molecular virial coefficients as will be shown below. A
key feature is its dependence on the system volume. Specifically,
although it has the limiting behaviour $\lim_{V\to\infty} f_N(V)=1$,
for finite system volume $f_N(V)$ deviates from unity.  However, on the
face of it, determining $f_N(V)$ by simulation via eq.~\ref{eq:limit}
is not a feasible proposition since it entails populating a
$3N$-dimensional histogram for $P_N(\br^N)$ with sufficient
statistics to yield precise probabilities. Fortunately, though, it
turns out to be possible to determine $f_N(V)$ using only
one-dimensional histograms. To see this, consider the quantity
\be
g^\prime_N(r_{\rm min})\equiv\frac{P_N(r_{\rm min})}{P_N^{\rm ig}(r_{\rm min})}\:.
\label{eq:gNrmin}
\ee
Here $r_{\rm min}$ is, for some configuration, the smallest, ie. the {\em minimum} 
separation among the $N$ tagged sites. In the course of a simulation, one
can accumulate histograms for $P_N(r_{\rm min})$ and $P_N^{\rm ig}(r_{\rm
  min})$ and thus form $g^\prime_N(r_{\rm min})$. Clearly, though,
the limit $r_{\rm min}\to\infty$ is none other than the
limit $|\br_k-\br_l|\to\infty, \:\;\forall\; k,l$. Moreover, since in
this limit the microstates of the tagged particles are visited with
constant probability $\Omega^N Z_N^{-1}$, while those of the ideal gas are
visited with probability $V^{-N}$, it follows that the
limiting value of $g^\prime_N(r_{\rm min})$ is the same as that of
$\tilde{g}_N(\br^N)$, i.e.
\be
\lim_{r_{\rm min}\to \infty}g^\prime_N(r_{\rm min})= f_N(V)\:.
\label{eq:gNrminlim}
\ee

Eq.~\ref{eq:gNrminlim} provides a straightforward computational prescription for
determining $f_N(V)$, which in turn permits the calculation of the
virial coefficients for the molecular system. Specifically, 
from the virial cluster expansion \cite{Hill:1988ye} one finds that for $N=2$ particles
\be
B_2= \frac{V}{2}\left(1-\frac{Z_2}{Z_1^2}\right)    = \frac{V}{2}\left( 1-\frac{1}{f_2(V)} \right )\:.
\label{eq:B2}
\ee
Similarly for three particles one finds 
\bea
B_3 &=& \frac{V^2(Z_1^4-3Z_2Z_1^2-Z_3Z_1+3Z_2^2)}{3Z_1^4}\nonumber\\
    &=& 4B_2^2-2B_2V+V^2\frac{(f_3(V)-1)}{3f_3(V)}\:.
\label{eq:B3}
\eea
More generally, knowledge of $f_\gamma(V), \gamma=2,\ldots, n$ permits the
calculation of the virial coefficient $B_n$.  

Thus measurements (for a small number of molecules) of the asymptotic value of a simple-to-measure structural
quantity, $g^\prime_N(r_{\rm min})$, provide direct access to
molecular virial coefficients. The utility of the approach is wide
because it can be used in conjunction with any simulation method
capable of producing equilibrium configurations, for example Molecular
Dynamics (MD), Monte Carlo (MC) or Langevin Dynamics.  Furthermore it
can deal with much more complex systems than is possible with an
existing method~\cite{Singh:2004zr}.

In general one can estimate $f$ visually, or from a
fit. However, we have found that a particularly accurate measure
results from the ratio of integrals
\be
f_N(V)=\frac{\int_{r_l}^{r_u} P_N(r_{\rm min}){\rm d}r_{\rm min}}{\int_{r_l}^{r_u} P^{\rm ig}_N(r_{\rm min}){\rm d}r_{\rm min}}\:,
\label{eq:findf}
\ee where $r_l$ is some value of $r_{\rm min}$ for which
$g^\prime(r_{\rm min})$ can be considered to have first reached its
limiting value, and $r_u$ is the largest value of $r_{\rm min}$ for
which data has been accumulated, which will typically be half the
simulation box diagonal length. It should be emphasized that in
practice eq.~(\ref{eq:findf}) is evaluated simply from a count of
entries in the respective histograms for $P_N(r_{\rm min})$ and
$P_N^{\rm ig}(r_{\rm min})$--no numerical quadrature is necessary.  

\begin{table}[h]
\begin{tabular}{cccc}
$N$ & $V$ & $B_N$ & $B_N^{\rm exact}$\protect\cite{Clisby:2004lh}\\\hline
2 & $(2.5\sigma)^3$ & $2.09441(6)\sigma^3$ & $2.0943951\ldots\sigma^3$\\
3 & $(3.5\sigma)^3$ & $2.7418(4)\sigma^6$ &  $2.7415567\ldots\sigma^6$ \\
4 & $(3.5\sigma)^3$ & $2.629(22)\sigma^9$ &    $2.6362180\ldots\sigma^9$ \\\hline
\end{tabular}
\caption{Estimates of the first four virial coefficients of hard spheres, compared with exact values.}
\label{tab:HS}
\end{table}

To test the method we have used it to estimate the first few virial
coefficients of a single component system of hard spheres of diameter
$\sigma$, finding excellent agreement with exact values (see
Tab.~\ref{tab:HS}).  Having validated the method on a simple system,
we turn to a more challenging problem, namely that of quantifying the
scale of many-body effects in CG models for colloid-polymer
mixtures. In such systems the polymers mediate effective colloidal
interactions~\cite{Lekkerkerker:2011}. A commonly studied model treats
the colloids as big hard spheres of diameter $\sigma_b$, and the
polymers as small hard spheres of diameter $\sigma_s$, so that the
size ratio is $q\equiv\sigma_s/\sigma_b$.  The effective Hamiltonian
then provides a CG description of the colloidal interactions in which
the polymer degrees of freedom have been integrated out. Quite
generally it takes the form $H^{\rm eff}=H^0+\Theta$ where $H^0$ is
the bare colloid-colloid interaction, while $\Theta$ is a many-body
contribution arising from the polymers, which can in turn be written
as a sum over $n$-body terms
$\Theta=\sum_{n=1}^\infty\theta_n$~\cite{Dijkstra1999}.  Common
practice is to approximate this Hamiltonian in terms of a sum over
pair interactions, i.e.  $H^{\rm
  eff}\approx\sum_{i,j}[\phi(r_{ij})+W(r_{ij})]$ where $\phi(r_{ij})$
is the hard sphere interaction between a pair of colloids whose
centers are separated by a distance $r_{ij}$, while $W(r_{ij})$ is the
depletion pair potential, whose form depends on the small particle
volume fraction and model details such as the degree of additivity of
the big-small interaction.  Usually one assumes that the small
particles occupy an open ensemble, so that $W(r)$ is parameterized in
terms of the reservoir volume fraction $\eta_s^r$.

Since the depletion pair potential plays a central role in theories and
simulations of colloid-polymer mixtures, it is desirable to
quantify the effects of neglecting higher order terms in $H^{\rm eff}$, the most
prominent of which is triplet interactions. Our strategy for doing so
estimates the third virial coefficient $B_3^{\rm eff}$ for the full
effective fluid and compares it with the corresponding value $B_3^{\rm
  dep}$ for three particles interacting via the depletion pair
potential. This comparison directly probes the extent to which the
interaction between a pair of big particles is influenced by the proximity of a third one.

To obtain estimates for $B_3^{\rm eff}$ we deploy the geometrical
cluster algorithm (GCA)~\cite{Dress1995,Liu2004}. This efficient
rejection-free Monte Carlo scheme can generate equilibrium
configurations at practically any value of $q$.
We have used it to study systems of
$N=2$ and $N=3$ big particles in a sea of small ones at
various $\eta_s^r$. The procedure is as follows:

\begin{enumerate}

\item[(i)] In a simulation of $N=2$ tagged big particles, measure the
  form of $g^\prime_2(r)$ at some prescribed $\eta_s^r$. This yields
  the value of $B_2^{\rm eff}(\eta_s^r)$ via eq.~(\ref{eq:B2}).
\vspace*{-1mm}
\item[(ii)] Use the form of $g^\prime_2(r)$ obtained in (i) 
  to estimate the depletion potential $W(r|\eta_s^r)$ by employing 
  the procedure detailed by Ashton {\em et al}~\cite{Ashton:2011kx}.
\vspace*{-1mm}
\item[(iii)] Next simulate $N=3$ tagged big particles at the same value of
  $\eta_s^r$ and measure $\tilde{g}_3(r_{\rm min})$. Together
  with the estimate of $B_2^{\rm eff}(\eta_s^r)$ obtained in (i), this yields
  an estimate for $B_3^{\rm eff}(\eta_s^r)$ via eq.~(\ref{eq:B3}).
\vspace*{-1mm}
\item[(iv)] Finally perform a simple MC simulation of three particles
  interacting via the depletion potential $W(r|\eta_s^r)$
  obtained in (ii).  This yields the third virial coefficient $B_3^{\rm
    dep}(\eta_s^r)$ via eq.~(\ref{eq:B3}).

\end{enumerate}

We have applied this procedure to study two models of colloid-polymer
mixtures, namely the Asakura-Oosawa (AO) model and a system of
additive hard spheres.  The AO model describes colloidal hard-spheres
in a solvent of ideal polymer that have a hard-particle interaction
with the colloids~\cite{Asakura1954,Asakura:1958uq},
c.f. Fig.~\ref{fig:illustr}(a). Owing to its extreme non-additivity,
the exact form of the depletion potential is known analytically~\cite{Asakura:1958uq} which obviates the need to perform steps (i) and
(ii) above. Furthermore many-body forces are known to vanish for size
ratios $q<0.1547$~\cite{Brader2003,Dijkstra:1999kl}, a fact that allows us to
further test our methodology and its sensitivity.

\begin{figure}[h]
\includegraphics[type=pdf,ext=.pdf,read=.pdf,width=0.9\columnwidth,clip=true]{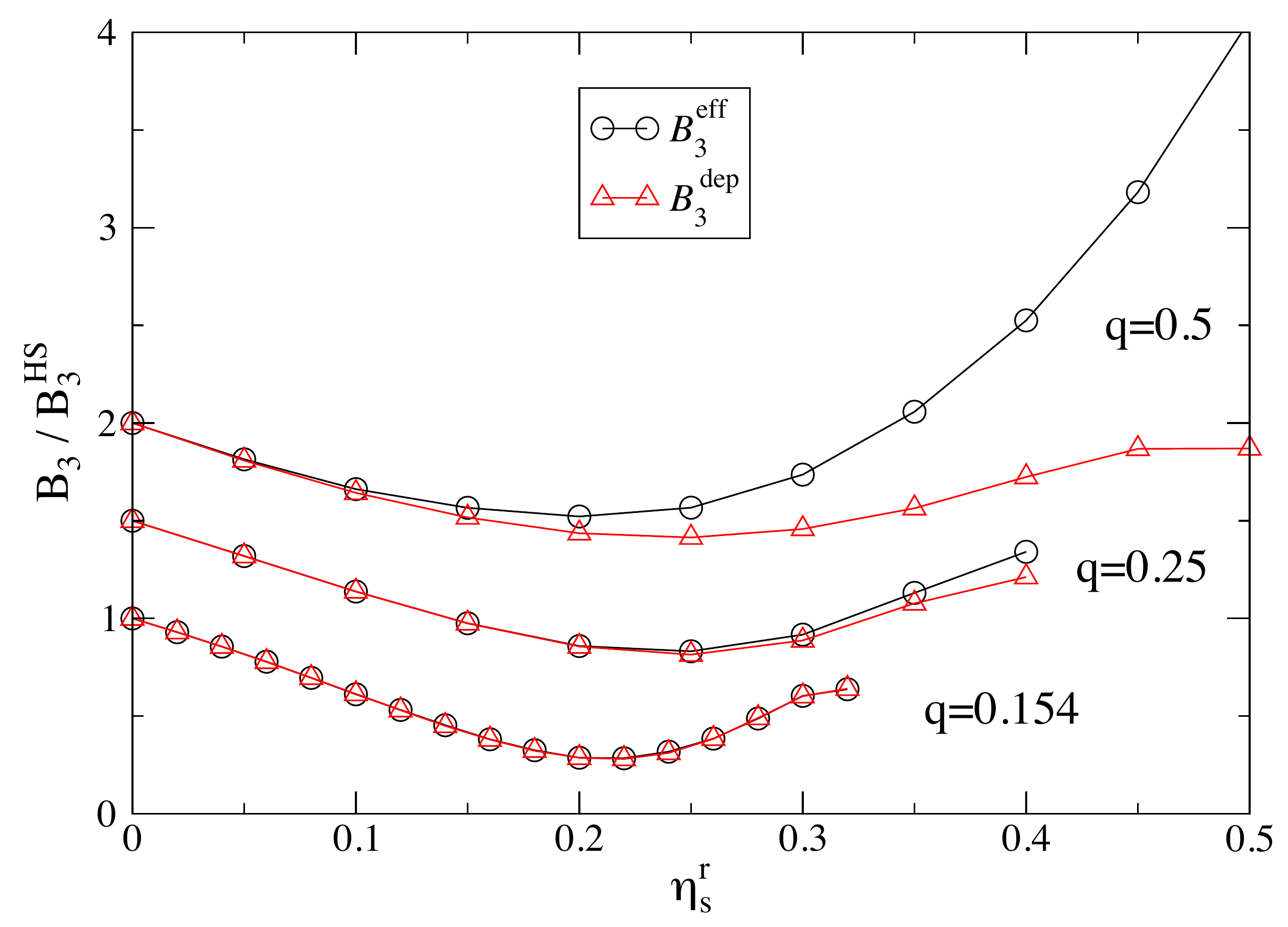}
\caption{Comparison of $B_3^{\rm
    eff}(\eta_s^r)$, and $B_3^{\rm dep}(\eta_s^r)$ for the AO
  model with size ratios $q=0.5, 0.25, 0.154$. Lines are guides to the eye and statistical uncertainties are
  smaller than the symbol sizes. To aid visibility, the curves for
  $q=0.25$ and $q=0.5$ have been shifted vertically by $0.5$ and $1.0$
  respectively.}
 \label{fig:B3ao}
 \end{figure}

Fig.~\ref{fig:illustr}(a) includes a sample plot of $g_3^\prime(r_{\rm
  min})$ obtained for the AO model using a box of size
$V=(3.5\sigma_l)^3$ at $\eta_s^r=0.2,q=0.154$. The data show the
approach to the asymptote, $f_3(V)$. From plots such as this we have
obtained estimates of $B_3^{\rm eff}$ and $B_3^{\rm dep}$ for size
ratios $q=0.5,0.25, 0.154$, as shown in Fig.~\ref{fig:B3ao}. One
expects that triplet interactions, as quantified by the difference between
$B_3^{\rm eff}(\eta_s^r)$ and $B_3^{\rm dep}(\eta_s^r)$, should
increase with $\eta_s^r$ and this is indeed the case. We find that
$B_3^{\rm eff}>B_3^{\rm dep}$, consistent with the fact that triplet
interactions weaken the attraction between particles~\cite{Goulding:2001fk}. One further expects $B_3^{\rm eff}-B_3^{\rm
  dep}$ to diminish with decreasing $q$ and have vanished by
$q=0.154$, a feature which is confirmed to high precision by our data.

For additive hard spheres, the GCA is considerably less efficient than for the AO
model being limited to $\eta_s^r\lesssim 0.2$.  Although
triplet interactions are always present in principle, our
results (not shown) indicate that within this more limited 
range of $\eta_s^r$, they are negligibly small for $q=0.2$ and
$q=0.1$. This finding suggests that for applications at low to
moderate $\eta_s^r$ and small $q$ it is safe to use depletion potentials for
additive hard spheres.

As a final illustration of the power and generality of our method, we
have used it to quantify the role of triplet interactions in a model
for star polymers in implicit solvent, c.f. Fig.~\ref{fig:illustr}(b). Each star comprises a core particle
to which are attached a number (called the ``functionality'')
of linear polymer chains each comprising $n$ monomers. Bonded monomers
interact via a FENE spring, while non bonded monomers experience a
Lennard-Jones (LJ) potential. Using MD we have studied various combination
of functionality and chain length $n$. Our aim was to determine how these
parameters affect the size of the triplet interactions. In order to effect this comparison in a fair
manner, we tuned the temperature in each case such that $B_2$ matches a
prescribed value, thereby providing a ``corresponding state''. The procedure for measuring the size of
triplet interactions via virial coefficients is similar to that
outlined for the colloid-polymer mixtures, except that the tagged
particles are now taken to be the set of core atoms. The pair
potential is the potential of mean force (pmf) which is obtained in a
simulation of two stars. We then simulate three particles interacting
via this potential to obtain $B_3^{\rm pmf}$. This we compare with
  $B_3^{\rm star}$, measured in a simulation of $N=3$ star polymers (a sample plot of
$g_3^\prime(r_{\rm min})$ in a box of volume $V=(40\sigma)^3$
is included in Fig.~\ref{fig:illustr}(b)). 

\begin{figure}[h]
\includegraphics[type=pdf,ext=.pdf,read=.pdf,width=0.9\columnwidth,clip=true]{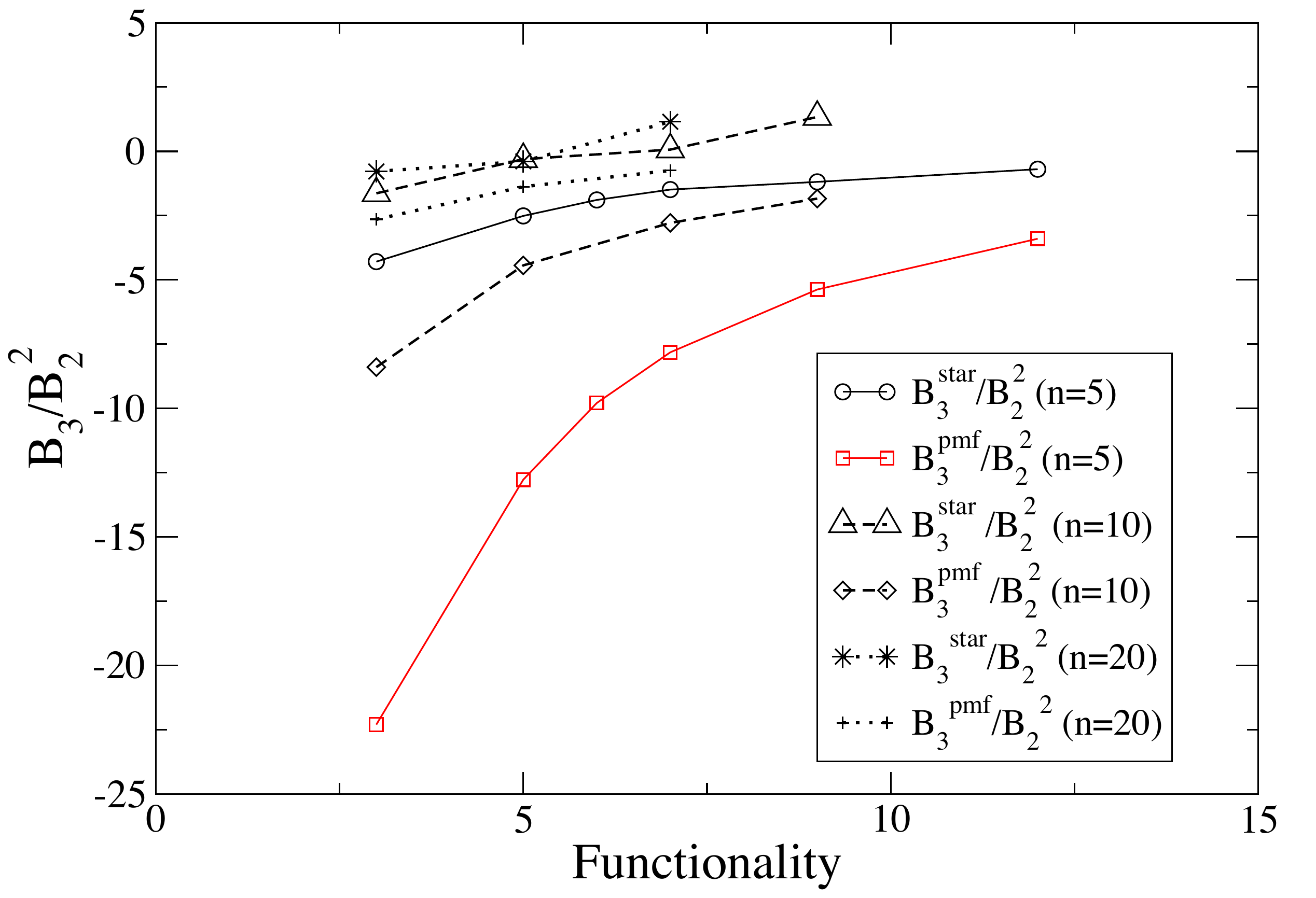}
\caption{Estimates of the dimensionless third virials $B_3^{\rm pmf}/B_2^2$ and $B_3^{\rm star}/B_2^2$ vs
  functionality for various chain lengths $n$. Volumes ranged
from $V=(20\sigma)^3$ to $V=(40\sigma)^3$, large enough to
  access the limiting behaviour of $g_3^\prime(r_{\rm min})$. Bonded
  monomers interact via a FENE potential with parameters
  $K=30.0\epsilon/\sigma^2, R_0=1.5\sigma$~\cite{Kremer:1990rt}. The
  LJ potential was truncated and shifted at $r=2.5\sigma$. In all cases
 $T$ is chosen to yield $B_2=-3321\sigma^3$. Errors are comparable with symbol sizes.}
 \label{fig:star_results}
 \end{figure}

The results are shown in Fig.~\ref{fig:star_results} and reveal large
discrepancies between $B_3^{\rm pmf}$ and $B_3^{\rm star}$, which
decrease in magnitude as both the functionality and the arm length
increase. Clearly the disparity is such that one should expect a quite
different equation of state (as well as other thermodynamic
quantities) to arise from the coarse-grained system described by the
pmf compared to the full model. We believe that the importance of
many-body effects in this system arises from the ability of the
polymers to substantially overlap, which occurs predominantly for
lower functionality and smaller number of monomers per arm. When two
polymers overlap, the resulting composite particle is locally much
denser than for a single polymer. Accordingly a third polymer is much
less likely to overlap with the first two due to short ranged
monomeric repulsions. Clearly, however, this effect is completely
neglected in the pair potential framework. This observation should be
relevant to CG models for many other types of polymer-based soft
particles, including cluster forming amphiphilic dendrimers~\cite{Lenz:2012fk}.

In summary, we have proposed a general method for calculating low-order
virial coefficients of complex fluids via a simple-to-measure
structural property. The method is the only direct approach (of which
we are aware) for achieving this. We have highlighted its utility in
quantifying the consequences of neglecting many-body effects in
coarse-graining schemes. Beyond this it should prove useful as a means
of testing new and existing molecular models by comparing the extent to which they
reproduce experimentally determined virial coefficients.

\acknowledgments

This work was supported by EPSRC grants EP/F047800 and
EP/I036192. We thank Bob Evans, Rob Jack, Andrew
Masters and Friederike Schmid for helpful discussions.

%

\end{document}